\documentclass[format=sigconf]{acmart}
\renewcommand\footnotetextcopyrightpermission[1]{}
\usepackage{fancyhdr}
\begin{document}


\title{Learning-Based Video Coding with Joint Deep Compression and Enhancement}


%
%
\author{Tiesong Zhao}
\email{t.zhao@fzu.edu.cn}
\affiliation{%
 \institution{Fuzhou University}
 \country{ }}
%

\author{Weize Feng}
\email{201127019@fzu.edu.cn}
\affiliation{%
 \institution{Fuzhou University}
 \country{ }}

\author{Hongji Zeng}
\email{201120063@fzu.edu.cn}
\affiliation{%
 \institution{Fuzhou University}
 \country{ }}

\author{Yuzhen Niu}
\email{yuzhenniu@gmail.com}
\affiliation{%
 \institution{Fuzhou University}
 \country{ }}

\author{Jiaying Liu}
\email{liujiaying@pku.edu.cn}
\affiliation{%
 \institution{Peking University}
 \country{ }}

%
%
%
%



\begin{abstract}
The end-to-end learning-based video compression has attracted substantial attentions by paving another way to compress video signals as stacked visual features. This paper proposes an efficient end-to-end deep video codec with jointly optimized compression and enhancement modules (JCEVC). First, we propose a dual-path generative adversarial network (DPEG) to reconstruct video details after compression. An $\alpha$-path facilitates the structure information reconstruction with a large receptive field and multi-frame references, while a $\beta$-path facilitates the reconstruction of local textures. Both paths are fused and co-trained within a generative-adversarial process. Second, we reuse the DPEG network in both motion compensation and quality enhancement modules, which are further combined with other necessary modules to formulate our JCEVC framework. Third, we employ a joint training of deep video compression and enhancement that further improves the rate-distortion (RD) performance of compression. Compared with x265 LDP very fast mode, our JCEVC reduces the average bit-per-pixel (bpp) by 39.39\%/54.92\% at the same PSNR/MS-SSIM, which outperforms the state-of-the-art deep video codecs by a considerable margin.
\end{abstract}

\keywords{video coding, deep video compression, end-to-end video codec}


\maketitle

\section{Introduction}

\label{sec:intro}

High efficient video compression has been a challenging task in multimedia community since 1980s. During the past four decades, the researchers have devoted to improve the rate-distortion (RD) efficiency by introducing more coding tools, such as hierarchical predictions, coding tree units and asymmetric partitions, to the hybrid video coding structure. In each generation of video codec, these consisting efforts approximately halves the compressed bits at the same visual quality. Among them, the popular H.265/HEVC \cite{Sullivan:TCSVT12} and H.266/VVC \cite{Bross:TCSVT21} are considered as the newest achievements of Joint Collaborative Team on Video Coding (JCT-VC). With the widespread use of high definition (HD) videos, it is undoubtable that the video coding problem is still a critical issue in the 5G or B5G era.
\begin{figure}[htbp]
	\centering
	\includegraphics[width=1.0\linewidth]{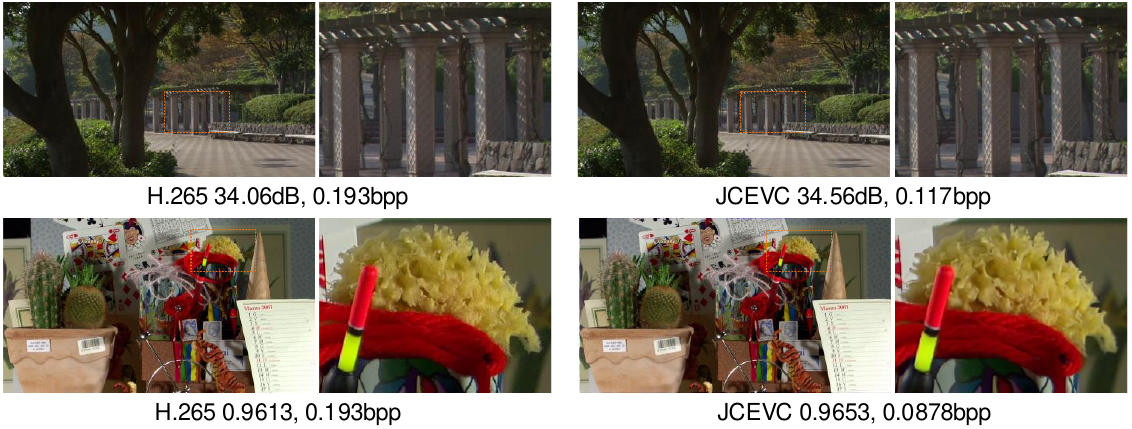}
	\caption{Reconstructed frames with the H.265 (x265 LDP very fast) and our JCEVC. The JCEVC reduces the bit-per-pixel (bpp) of H.265 by almost half whilst retaining competitive PSNR or MS-SSIM. In the view of video coding, the JCEVC achieves a significantly higher RD performance.}
	\label{fig:example}
\end{figure}
The popular video codecs treat the videos as {\em signals}. They remove the spatial-temporal redundancies of videos by low-level transform, quantization and entropy coding. However, in computer vision community, the videos can be processed as stacked {\em features}, which allows us to develop end-to-end video codecs with big data and learning. Recently, the learning-based image codecs \cite{CaiC:TIP20, CaiJ:TIP20, ChenLH:TIP21, ChenT:TIP21, ChengZ:CVPR20, HeD:CVPR21, Johnston:CVPR18, LeeJ:ICLR19, MaH:TPAMI20, YangF:CVPR21} have surpassed the traditional image codecs in terms of compression efficiency, which also inspired learning-based codecs for videos. In fact, we have witnessed a booming of learning-based video codecs in the past two years.

The learning-based video codecs utilize the deep neural networks to imitate the motion estimation (ME), motion compensation (MC), residual compression and video reconstruction \cite{Agustsson:CVPR20, Habibian:ICCV19, HuZ:ECCV20, HuZ:CVPR21, Klopp:ICCV21, LinJ:CVPR20, LiuB:CVPR21, LuG:CVPR19, LuG:ECCV20, LuG:TPAMI21, Rippel:ICCV19, Rippel:ICCV21, YangR:CVPR20, YangR:JSTSP21}. Owing the advantage of deep learning on large-scale datasets, these methods adopt convolutional neural networks (CNNs), auto-encoders, and/or generative adversarial networks (GANs) to achieve the end-to-end video compression. Among them, \cite{YangR:CVPR20, YangR:JSTSP21} utilized bi-directional prediction while the other methods used one-way prediction with 1 to 4 references. A separate quality enhancement network was deployed in post-processing stage of \cite{YangR:CVPR20}. Early works exhibited superior performances than H.264/AVC or competitive performances compared with H.265/HEVC \cite{Habibian:ICCV19, LuG:CVPR19, Rippel:ICCV19, Agustsson:CVPR20}. Recently, the deep video codecs have surpassed H.265/HEVC in terms of RD performance \cite{HuZ:ECCV20, HuZ:CVPR21, Klopp:ICCV21, LinJ:CVPR20, LiuB:CVPR21, LuG:ECCV20, LuG:TPAMI21, Rippel:ICCV21, YangR:CVPR20, YangR:JSTSP21}. In such sense, the end-to-end learning-based video codecs have paved another way of video compression.

Despite of these great efforts, the RD performance of end-to-end deep video coding is still inferior to the newest reigning video codec, H.266/VVC. It is imperative to further improve the RD efficiency of deep video codecs. In this paper, we move the next step to propose an efficient deep video codec that benefits from GAN-based visual enhancement. To improve the full-reference reconstruction quality with GAN, we introduce a parallel path with residual attention blocks (RABs). This dual-path enhancement with GAN (DPEG) network is co-trained by a generative-adversarial process to well reconstruct the video frames after quantization, where a convolutional long short-term memory (ConvLSTM) network \cite{ConvLSTM} is also incorporated to refer to multiple coded frames. In our codec, this design is reused in both MC and perceptual enhancement, while the optical flow, auto-encoders and residual networks are utilized to construct the other modules. Aiming at an optimal RD performance, we employ a joint training of deep video compression and enhancement. The proposed joint compression and enhancement for deep video coding (JCEVC) achieves superior performance compared with H.265, as depicted in Figure \ref{fig:example}.

Our main contributions are summarized as follows.

\textbf{A DPEG network for video reconstruction}: We propose the DPEG with two paths of different receptive fields. An $\alpha$-path focuses on the structure features with auto-encoder and ConvLSTM. A $\beta$-path focuses on the texture details with RABs. The fusion of these features improves the visual quality of video frames and also reduces the bpp for residual coding.

\textbf{A JCEVC framework for end-to-end deep video coding}: We propose the JCEVC framework by reusing DPEG network in both MC and perceptual enhancement. The other modules of our JCEVC are constructed by CNN-based optical flow, auto-encoder and residual networks.

\textbf{Joint training of video compression and enhancement}: We employ a joint training of compression and enhancement in JCEVC framework, in order to achieve an optimal tradeoff between compressed bits and reconstructed quality. To the best of our knowledge, we are the first to jointly optimize compression and enhancement in end-to-end deep video coding. Experimental results reveal the effectiveness of our JCEVC with joint training.

\section{Related Work}
\label{sec:ref}

\textbf{Deep image compression.} The reigning image codecs, such as JPEG \cite{JPEG}, JPEG2000 \cite{JPEG2000} and BPG \cite{BPG}, employ the frequency transform and quantization to remove the spatial redundancies of images. While in deep image codecs, the auto-encoders, recurrent neural networks (RNNs) and GANs are widely used. Recent efforts have supported spatial rate allocation, {\it i.e.}, to allocate bits based on the spatial textures and contexts \cite{CaiC:TIP20, ChenT:TIP21, ChengZ:CVPR20, Johnston:CVPR18, LeeJ:ICLR19} or multiple bpps with one network \cite{CaiJ:TIP20, YangF:CVPR21}. In \cite{ChenLH:TIP21}, a CNN-based ProxIQA model was proposed to mimic the perceptual model for RD tradeoff. In \cite{ChengZ:CVPR20}, the discretized Gaussian mixture likelihoods were utilized to parameterize the latent code distributions, aiming at a more accurate and flexible entropy model. In \cite{HeD:CVPR21}, a checkerboard context model was proposed to support parallel image decoding. In \cite{MaH:TPAMI20}, the CNN was employed to design a wavelet-like transform for removing redundancies. These methods exploits the spatial correlations of single pictures, while our JCEVC framework focuses on the spatial-temporal correlations of successive pictures, especially the MC module based on DPEG.

\textbf{Deep video compression.} By removing the spatial-temporal redundancies of video frames, the reigning video codecs, such as H.265/HEVC and H.266/VVC, achieve significantly higher compression efficiency than those image codecs. These characteristics have also been utilized in deep video coding. A classic method, called deep video compression (DVC) \cite{LuG:CVPR19}, replicated the ME/MC, transform/quantization and entropy coding with optical flow, non-linear residual encoder and CNN, respectively. In \cite{LuG:ECCV20}, an error-propagation-aware training was proposed to address the error propagation and content adaptive compression in DVC. In \cite{LuG:TPAMI21}, two variants of DVC, DVC Lite and DVC Pro, were designed with different coding complexities. In \cite{LinJ:CVPR20}, the CNN-based motion vector (MV) prediction, MV refinement, multi-frame MC and residual refinement were introduced to develop the multiple frames prediction for learned video compression (M-LVC). In \cite{YangR:CVPR20}, hierarchical learned video compression (HLVC) introduced the hierarchical group-of-picture (GOP) structure which had shown its high efficiency since H.264 scalable coding. It also utilized a weighted recurrent quality enhancement to further improve the visual quality at decoder-end. In \cite{YangR:JSTSP21}, a recurrent learned video compression (RLVC) employed the recurrent auto-encoder and recurrent probability model for improved MV and residual compression.

The rate allocation of deep video coding was first realized by \cite{Habibian:ICCV19} and \cite{Rippel:ICCV19}, which utilized the deep generative model and recursive network for high compression efficiency, respectively. \cite{Agustsson:CVPR20} designed a scale-space flow to improve the ME robustness under common failure cases, {\it e.g.}, disocclusion and fast motion. \cite{HuZ:ECCV20} designed a resolution-adaptive flow coding (RaFC) to effectively compress the optical flow maps globally and locally. In \cite{Klopp:ICCV21}, the encoder complexity is optimized with less model parameters. In \cite{Rippel:ICCV21}, an efficient, learned and flexible video coding (ELF-VC) was proposed to support flexible rate coding with high RD efficiency. Recently, \cite{HuZ:CVPR21} and \cite{LiuB:CVPR21} came up with different ways (CNN or GAN) to compress video contents via low-dimensional feature representations, which were further utilized to reconstruct video frames. In our work, the dual-path DPEG network is introduced to extensively reduce the spatial-temporal redundancies of video frames. Compared with the state-of-the-arts, our method shows a significantly increased compression efficiency.

\textbf{Visual enhancement.} The visual enhancement techniques are utilized to improve the visual quality of images or videos. Generally, the image enhancement can be achieved by GAN \cite{ChenY:CVPR18, NiZ:TIP20} or CNN \cite{LiC:TPAMI21, SonT:ECCV20}. In \cite{YangR:CVPR18}, a multi-frame quality enhancement (MFQE) method utilized high-quality frames to enhance low-quality frames, where the high-quality frames were detected with support vector machine (SVM). In \cite{GuanZ:TPAMI21}, the MFQE2.0 method replaced the SVM by bi-directional long short-term memory (BiLSTM) and also improved the convolutional network of MFQE. \cite{Vimeo90k} proposed a task-oriented flow (TOFlow) that demonstrated its superiority to optical flow in video enhancement. \cite{Haris:CVPR20} leveraged the spatial-temporal relationship of videos and proposed to simultaneously increase their spatial resolutions and frame rates. The visual enhancement is also introduced at the decoder-end of HLVC \cite{YangR:CVPR20}. In our work, we incorporate the enhancement module into the reconstruction process, thereby leading to a joint training of video compression and enhancement.

\begin{figure*}[htbp]
	\centering
	\includegraphics[width=1.0\linewidth]{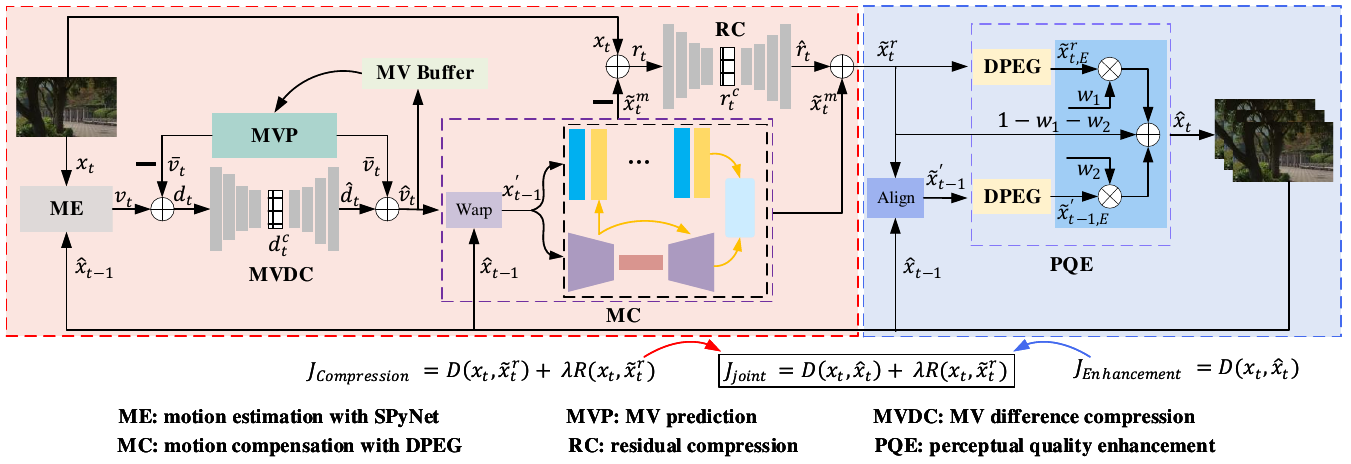}
	\caption{The framework of our JCEVC encoder, where both MC and PQE modules are realized by our DPEG network. For a frame $x_t$, the ME module calculates its MV $v_t$ with SPyNet and bi-directional IPPP structure. Meanwhile, the MVP module derives an MV prediction, $\overline v_t$, based on an MV buffer of references. The MVDC module compresses and reconstructs the MV difference $d_t=v_t-\overline v_t$ with an auto-encoder, quantization and entropy coding. Then, the MC module utilizes the reconstructed MV, $\hat v_t=\overline v_t+\hat d_t$, to align the reconstructed frame $\hat x_{t-1}$ to $x_t$. The warped frame $x_{t-1}^{'}$ is enhanced by the DPEG network to generate a compensated frame $\tilde x_t^m$. After that, the RC module compresses and reconstructs the texture residual $r_t=x_t-\tilde x_t^m$. Finally, the PQE module receives the residual compensation $\tilde x_t^r=\tilde x_t^m+\hat r_t$ and the aligned reference $\tilde x_{t-1}^{'}$, and employs two DPEGs with weighted fusion to obtain the reconstructed frame $\hat x_t$. Aiming at an optimal RD efficiency, the compression and enhancement processes are jointly trained.}
	\label{fig:framework}
\end{figure*}

\section{Proposed Method}
\label{sec:method}

\textbf{Notations.} Let $x_t$ denote the $t$-th frame in picture coding order of an original video and $\hat x_t$ denote its constructed frame. $\tilde x_t^m$ and $\tilde x_t^r$ represent the compensated frames of $x_t$ with motion and residual information, respectively. The reconstructed $(t-1)$-th frame, $\hat x_{t-1}$, is sent back for recurrent prediction. It is also aligned to $x_t$ and $\tilde x_t^r$, resulting to $x_{t-1}^{'}$ and $\tilde x_{t-1}^{'}$, for MC and quality enhancement. The MV matrix of $x_t$ is predicted as $\overline v_t$ and finally denoted as $v_t$ after ME. Their difference, $d_t=v_t-\overline v_t$, is compressed and reconstructed as $\hat d_t$. The difference between $x_t$ and $\tilde x_t^m$ is represented as a residual $r_t$, whose corresponding reconstruction is denoted by $\hat r_t$.

\subsection{The JCEVC framework}
\label{subsec:framework}

As shown in Figure \ref{fig:framework}, the encoder of JCEVC consists of 6 modules: ME, MVP, MVDC, MC, RC and PQE, among which we redesign the MC and PQE modules and improve the remaining modules. All modules are jointly trained for an optimized RD performance. This coding procedure applies to all P frames while the context-adaptive entropy model of \cite{LeeJ:ICLR19} is utilized to compress I frames.

\textbf{ME module.} In JCEVC, we employ a bi-directional IPPP structure \cite{YangR:JSTSP21} with a GOP size of 15. The frames 0, 15 are coded as I frames while the others are coded as P frames.
Among them,
the frames 1$\sim$7 are forwardly predicted from frame 0 while the frames 8$\sim$14 are backwardly predicted from frame 15 of next GOP. To remove temporal redundancies, the ME module estimates the MV $v_t$ between adjacent frames: $x_t$ and $\hat x_{t-1}$. In this paper, we employ a low-complexity optical flow model, SPyNet \cite{SPyNet}, which combines spatial pyramid and deep convolutional network for fast ME.


\textbf{MVP module.} Due to high spatial-temporal correlations between MVs, it is sensible to compress the MV difference $d_t$ instead of MV values. We set $d_t = v_t-\overline v_t$, where $\overline v_t$ is a predicted MV from an MV buffer, which is constituted by the MV information of its three preceding frames. The prediction is achieved by a light network with a convolutional layer, two residual blocks and another two convolutional layers. The channel number is 2 for the last layer and 64 for each of the others. The convolutional kernel size and stride are 3$\times$3 and 1, respectively. Relu is utilized as the activation function in all convolutional layers.

\textbf{MVDC module.} The MV difference $d_t\in R^{H\times W \times 2}$, where $H$ and $W$ are frame height and width, is compressed by MVDC module. To reduce the coding bits, we employ an auto-encoder with four downsampling layers and four upsampling layers that are implemented with convolutions and deconvolutions, respectively. The compact representation $ d_t^c\in R^{\frac{H\times W}{16}\times 128}$ is further processed by quantization and entropy coding, with the procedure presented in \cite{LuG:CVPR19}. The reconstructed MV can be calculated as $\hat v_t=\overline v_t+\hat d_t$.

\textbf{MC module.} The MC module utilizes the reconstructed previous frame, $\hat x_{t-1}$ and the reconstructed MV, $\hat v_t$, to generate a warped frame that is aligned to the current frame $x_t$. The warped frame, namely $x_{t-1}^{'}$, is fed into the DPEG network to reconstruct an enhanced frame $\tilde x_t^m$. A ConvLSTM model is deployed to use multiple reference frames in history. This module will be elaborated in Section \ref{subsec:mc}.

\begin{figure*}[htbp]
	\centering
	\includegraphics[width=1.0\linewidth]{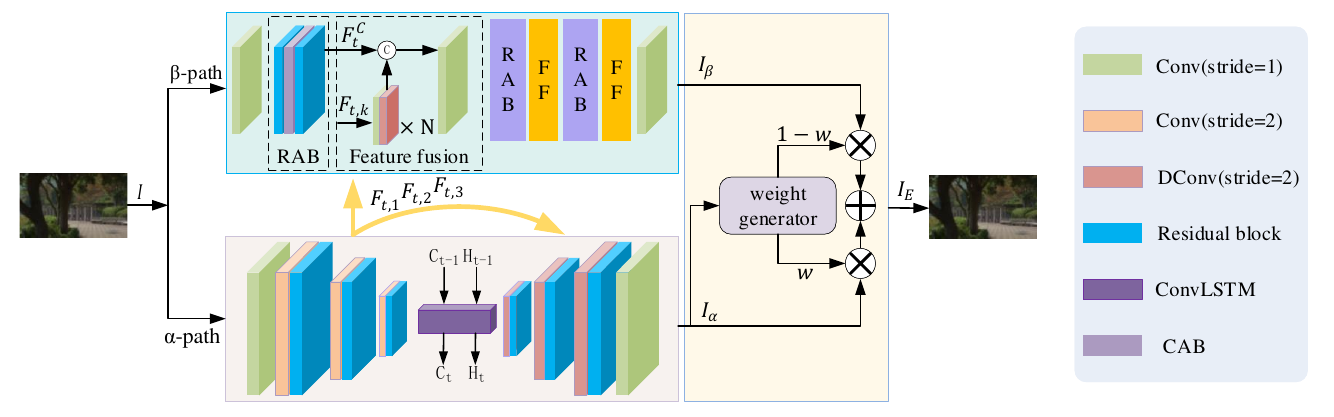}
	\caption{The proposed DPEG network. An $\alpha$-path with generator and ConvLSTM focuses on larger receptive field and global structures; a $\beta$-path with RABs focuses on smaller receptive field and local textures. Their outputs are combined with a weighted fusion. The sematic features ($F_{t,1}, F_{t,2}, F_{t,3}$) are fed from $\alpha$-path into $\beta$-path as a guidance. In particular, the input $I$ and output $I_E$ are task-dependent, {\it e.g.} $I=x_{t-1}^{'}, I_E=\tilde x_{t}^{m}$ in MC.}
	\label{fig:dpeg}
\end{figure*}

\textbf{RC module.} The motion compensated and enhanced frame $\tilde x_t^m$ is further utilized to calculate the texture residual $r_t=x_t-\tilde x_t^m$. Its compression is finished with the same process to that of MDVC. After this step, the reconstructed texture residual and compensated frame are represented by $\hat r_t$ and $\tilde x_t^r=\tilde x_t^m + \hat r_t$, respectively.

\textbf{PQE module.} The last module employs two DPEGs to enhance $\tilde x_t^r$ and $\tilde x_{t-1}^{'}$, and further combines them as the reconstructed frame $\hat x_t$. In particular, $\tilde x_{t-1}^{'}$ is a warped frame that aligns $\hat x_{t-1}$ to $\tilde x_t^m$, where the alignment process is realized by an ME and a warping operation. Details of this module will be presented in Section \ref{subsec:pqe}.

\textbf{The decoder.} The compressed stream of JCEVC consists of compressed MV residuals, compressed texture residuals and the information of previously coded frames, from which we can easily obtain $\hat d_t$, $\hat r_t$, $\hat x_{t-1}$ and $\overline v_t$. Then $\tilde x_t^m$ can be derived with $\hat v_t=\overline v_t+\hat d_t$, $\hat x_{t-1}$ and the MC module. Finally, the reconstructed frame $\hat x_t$ is obtained by $\tilde x_t^r = \tilde x_t^m + \hat r_t$, $\hat x_{t-1}$ and the PQE module.

\subsection{MC with DPEG network}
\label{subsec:mc}

In video coding, P frames generally have lower residuals than I frames. The residual at $t$-th frame, $r_t$ is calculated between the current frame $x_t$ and its motion compensated reference $\tilde x_t^m$. To further reduce the bits for $r_t$ under the same visual quality, we first compensate the previous frame with a non-linear warp,
\begin{equation}
	x_{t-1}^{'}={\rm Warp}(\hat{x}_{t-1},\hat{v}_{t} );
	\label{eq:mc1}
\end{equation}
and then enhance the warped frame with DPEG network,
\begin{equation}
	\tilde x_{t}^{m}={\rm DPEG}(x_{t-1}^{'}).
	\label{eq:mc2}
\end{equation}

The proposed DPEG network is implemented with dual-path GAN and RABs. To date, dual-path networks have been adopted in parallel processing of image denoising and enhancement tasks; however, they have not yet been exploited in image reconstruction after compression. The basic generator of GAN involves a de facto downsampling process to increase its receptive field, whilst excluding texture details in the original frame. This is unhelpful in frame reconstruction that is evaluated by full-reference quality metrics. To address this issue, we add a complementary path with RABs for video details, as shown in Figure \ref{fig:dpeg}. To avoid high computational complexity, the DPEG is designed as a compact framework with input $I$ and output $I_E$. In the second step of MC as Equation (\ref{eq:mc2}), $I=x_{t-1}^{'}, I_E=\tilde x_{t}^{m}$.

\textbf{$\alpha$-path.} To increase the receptive field of low-dimensional features, an encoder is employed with a convolutional layer with stride 1, three convolutional layers with stride 2 and three residual blocks. The obtained sematic features, $F_{t,1} \in R^{H \times W \times C}$, $F_{t,2} \in R^{\frac{H}{2} \times \frac{W}{2} \times C}$, $F_{t,3} \in R^{\frac{H}{4} \times \frac{W}{4} \times 2C}$, are fed into $\beta$-path and the decoder. After that, a ConvLSTM is inserted to fully utilize the reference information of coded frames. The state and output vectors of ConvLSTM, $C_{t-1}$ and $H_{t-1}$, are fed into current network. The decoder part is set as an inverse process of encoder with upsampling. To avoid sematic information loss, a U-Net \cite{UNet} is used to skip connect the sematic features to the decoder so that the enhanced results contain the original sematic information.

\textbf{$\beta$-path.} This path consists of two convolutional layers with stride 1, three RABs and three feature fusion blocks where downsamlping is not applied. An RAB is composed of two residual blocks and a channel attention block (CAB) \cite{CAB}. Each RAB extracts a feature representation $F_t^C \in R^{H \times W \times C}$ of $C$ channels, which is further fused with the sematic feature $F_{t,k}, k=1,2,3$ from $\alpha$-path. The feature fusion block also includes an upsampling process to match the dimensions of features. With a smaller receptive field, there is less access to global features in $\beta$-path. Therefore, the fusion of sematic features from $\alpha$-path benefits the frame reconstruction.

\textbf{Weighted fusion.} 
To take advantage of both paths, we employ a weighted summation of results:
\begin{equation}
	I_E = w \cdot I_{\alpha} + (1-w) \cdot I_{\beta},
	\label{eq:DPEG1}
\end{equation}
where $w$ is a weight matrix to represent the dependence degree of $I_E$ on $I_{\alpha}$. We utilize three convolutional layers and two residual blocks to extract the saliency of frame and further warp it to $(0,1)$ with a Sigmoid function.

\textbf{The discriminator.} The two paths of DPEG are co-trained by a generative-adversarial process after the weighted fusion of Equation (\ref{eq:DPEG1}). The training of this process involves a discriminator with attention mechanism. First, it utilizes four downsampling layers to achieve a larger receptive field. Then, it employs attention mechanism and different pooling strategies to generate two attention maps, $M_{avg}$ and $M_{max}$. After that, it multiplies the downsampled frame with the two attention maps and concatenates them as an feature map. Finally, it judges the obtained feature map after two convolutions with stride 1. The $M_{avg}$ and $M_{max}$ are also important in the loss function for training.


\subsection{PQE with DPEG networks}
\label{subsec:pqe}

The compensated video frames are further enhanced before reconstruction. There have been extensive studies to enhance the visual quality or remove compression artifacts of video sequences. The quality enhancement module has also been introduced to decoder-end of deep video compression by \cite{YangR:CVPR20}. In this paper, we deploy a PQE module before reconstruction, which is thus jointly trained by the encoder. The DPEG network is reused in this module due to its effectiveness in visual enhancement.


Aiming at a better visual quality, two parallel DPEG networks are adopted to enhance the compensated frame $\tilde x_t^r$ and the aligned previous frame $\tilde x_{t-1}^{'}$, respectively. The $\tilde x_t^r$ is compensated by $\tilde x_t^m$ and the reconstructed residual $\hat r_t$. The $\tilde x_{t-1}^{'}$ represents the results of aligning $\hat x_{t-1}$ to $\tilde x_t^r$, where the alignment process is a combination of ME and warping: an MV is firstly calculated by SPyNet and then utilized to warp $\hat x_{t-1}$:
\begin{equation}
	\tilde x_{t-1}^{'}={\rm Warp}(\hat x_{t-1},{\rm ME} (\hat x_{t-1}, \tilde x_t^r) ).
	\label{eq:pqe1}
\end{equation}
The $\tilde x_t^r$ and $\tilde x_{t-1}^{'}$ frames are separately enhanced as $\tilde x_{t,E}^r$ and $\tilde x_{t-1,E}^{'}$, which are fused to obtain the final reconstruction $\hat x_t$:
\begin{equation}
	\hat x_t = w_1 \cdot \tilde x_{t,E}^r + w_2 \cdot \tilde x_{t-1,E}^{'} + (1-w_1-w_2) \cdot \tilde x_t^r,
	\label{eq:pqe3}
\end{equation}
where the weights $w_1$ and $w_2$ are generated with the weight generator in Section \ref{subsec:mc} but are halved to avoid data overflow after summation. The weight $1-w_1-w_2$ is a penalty coefficient in case of enhancement failures.

\subsection{Joint training of JCEVC}
\label{subsec:training}

Our training process consists of two phases. In the first phase (0 $\sim$ 300K iterations), we perform a coarse-grain training with the reference frame $x_{t-1}$ and learning rate 1e-4; while in the second phase (300K $\sim$ 900 K iterations), we perform a fine-grain training with the reference frame $\hat x_{t-1}$ and learning rate 1e-5. In each phase, we successively train the MVP, MVDC, MC, RC, PQE modules and then perform a joint training of all modules. The ME module is performed with the SPyNet without further training.

The reasons to adopt this joint training strategy are as follows. In reigning video codecs, the RD optimization theory was proposed to minimize the RD cost of
\begin{equation}
	J_{\rm Compression} = D (x_t, \tilde x_t^r) + \lambda R (x_t, \tilde x_t^r),
	\label{eq:rdcost}
\end{equation}
where $D$ and $R$ refer the compression distortion and bitrate, separately. The objective of enhancement is to further reduce the distortion
\begin{equation}
	J_{\rm Enhancement} = D (x_t, \hat x_t) = D (x_t, {\rm PQE} (\tilde x_t^r, \hat x_{t-1})).
	\label{eq:enhdist}
\end{equation}
Compared with reigning video codecs, the deep video codec has an advantage that its end-to-end framework can be jointly optimized with training on a large-scale dataset. Taking this advantage, we propose to jointly train the compression and enhancement modules of deep video coding. The objective is then set as to minimize the total RD cost
\begin{equation}
	J_{\rm joint} = D (x_t, \hat x_t) + \lambda R (x_t, \tilde x_t^r).
	\label{eq:joint}
\end{equation}
Obviously, the joint training releases the burden of reconstruction. The compression stage allows a higher $D$, which can be eliminated in the
enhancement stage, with a reduced $R$. Hence, the overall RD tradeoff is improved.

Inspired by the RD cost function, we employ the following loss functions for the MVP, MVDC, MC, RC, PQE and joint training:
\begin{equation}
	\left\{
	\begin{array}{r@{\;=\;}l}
		\mathcal{L}_{\rm MVP} & {\rm MSE}(v_t,\overline{v}_{t})\\
		\mathcal{L}_{\rm MVDC} & {\rm MSE}(d_t,{\hat{d}}_t)+\lambda R_{mvd} \\
		\mathcal{L}_{\rm MC} & \mathcal{L}_\mathcal{G} \\
		\mathcal{L}_{\rm RC} & {\rm MSE}(r_t,{\hat{r}}_t)+\lambda R_{res}\\
		\mathcal{L}_{\rm PQE} & D(x_t,\hat{x}_t)\\
		\mathcal{L}_{\rm ALL} & D(x_t,{\hat{x}}_t)+\lambda(R_{mvd}+R_{res})
	\end{array}
	\right.,
	\label{eq:loss}
\end{equation}
where ${\rm MSE}$ denotes the mean squared error. $R_{mvd}$ and $R_{res}$ represent the bits consumed by MV difference and residuals after compression, respectively. $D(\cdot)$ represents the frame-level distortion, which is calculated as MSE and $1-$MS-SSIM in PSNR-oriented and MS-SSIM-oriented codecs, respectively. $\lambda$ is a coefficient for RD tradeoff. $\mathcal{L}_\mathcal{G}$ is the loss function for generator of DPEG. The loss functions for generator and discriminator are set as:
\begin{equation}
	\begin{split}
		\mathcal{L}_\mathcal{G}=\gamma E_{x\sim\hat{x}}[(x_t-\mathcal{G}(x_{t-1}^{'}))^2] \\
		+E_{x\sim\hat{x}}[1- \mathcal{D}(\mathcal{G}(x_{t-1}^{'}))^2]\\
		+E_{x\sim\hat{x}}[1-\phi (\mathcal{G}(x_{t-1}^{'}))^2],
	\end{split}
	\label{eq:ganloss1}
\end{equation}
\begin{equation}
	\mathcal{L}_\mathcal{D}=E_{x\sim\hat{x}}[1-\mathcal{D}(x_t)^2]+E_{x\sim\hat{x}}[\mathcal{D}(\mathcal{G}(x_{t-1}^{'}))^2],
	\label{eq:ganloss2}
\end{equation}
where $\phi(\cdot)$ represents the calculation of attention maps $M_{avg}$ and $M_{max}$.

\begin{figure*}[htbp]
	\centering
	\includegraphics[width=1.0\linewidth]{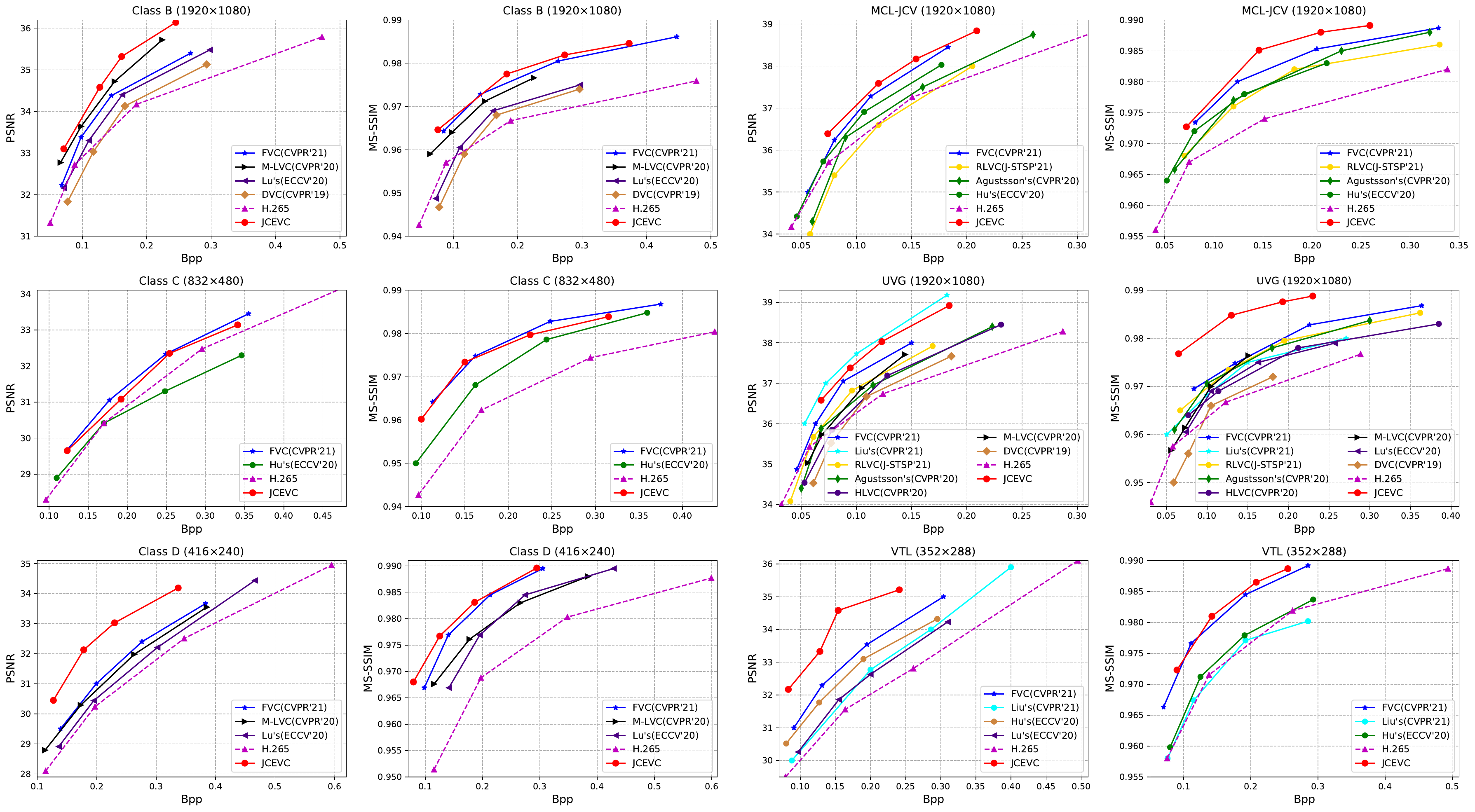}
	\caption{Compression of our JCEVC with 9 popular deep codecs and H.265 (x265 LDP very fast). The horizontal and vertical axes represent the coding bit (in bpp) and reconstructed quality (in PSNR or MS-SSIM), respectively. The proposed JCEVC exhibits significantly superior or at least competitive performance compared with the state-of-the-arts in each dataset.}
	\label{fig:curves}
\end{figure*}

\begin{table*}[htbp]
	\renewcommand\arraystretch{1.1}
	\resizebox{2.1\columnwidth}{!}{
		\begin{tabular}{c|c|c|c|c|c|c|c|c|c|c}
			
			\bottomrule
			\begin{tabular}[c]{@{}c@{}}Datasets\end{tabular} & DVC \cite{LuG:CVPR19}        & Hu's \cite{HuZ:ECCV20}           & Lu's \cite{LuG:ECCV20}         & Agustsson's \cite{Agustsson:CVPR20}    & HLVC \cite{YangR:CVPR20}         &M-LVC \cite{LinJ:CVPR20}         & RLVC \cite{YangR:JSTSP21}        & Liu's \cite{LiuB:CVPR21}   &FVC \cite{HuZ:CVPR21}          & \begin{tabular}[c]{@{}c@{}}JCEVC\end{tabular}          \\
			\hline
			Class B  & 5.66/-2.74 & \textbf{--}/\textbf{--}           & -13.35/-7.93 & \textbf{--}/\textbf{--}          & -11.75/-37.44 &  {\color{blue}-36.55}/-42.82 & -24.20/-50.42 &\textbf{--}/\textbf{--} & -23.75/ {\color{blue}-54.51} & {\color{red}-44.19} /{\color{red}-58.29} \\ 
			Class C  &25.88/-6.88    & 4.94/-32.44   & \textbf{--}/\textbf{--}          &  \textbf{--}/\textbf{--}            & 7.83/-23.63   &  \textbf{--}/\textbf{--}             & -4.67/-35.94& \textbf{--}/\textbf{--}  & {\color{red}-14.18} / {\color{blue}-43.58} &  {\color{blue}-8.58} /{\color{red}-44.10}  \\ 
			Class D  & 15.34/-18.51          & \textbf{--}/-32.43      & -6.86/\textbf{--}      &  \textbf{--}/\textbf{--}            & -12.57/ {\color{blue}-52.56} & -13.87/-36.27 &  {\color{blue}-27.01} /-48.85 & \textbf{--}/\textbf{--}& -18.39/-51.19 & {\color{red}-44.72} /{\color{red}-56.38} \\ 
			MCL-JCV &  \textbf{--}/\textbf{--}          & -10.60/-34.10 & 4.21/\textbf{--}       & -1.82/-33.61 &  \textbf{--}/\textbf{--}             &  \textbf{--}/\textbf{--}            &  \textbf{--}/\textbf{--}       & \textbf{--}/\textbf{--}      &  {\color{blue}-22.48} /{\color{red}-52.00} & {\color{red}-31.21} / {\color{blue}-51.17} \\ 
			UVG      & 10.40/8.05  & \textbf{--}/\textbf{--}             & -7.56/-25.49 & -8.80/-38.04 & -1.37/-30.12  & -12.11/-25.44 & -13.48/-40.62 &{\color{red}-49.42} /-30.70 & -28.71/ {\color{blue}-45.25} &   {\color{blue}{-47.62}}  / {\color{red}-77.60}  \\ 
			VTL      &  \textbf{--}/\textbf{--}          & \textbf{--}/-6.04       & -16.05/\textbf{--}     &  \textbf{--}/\textbf{--}            &  \textbf{--}/\textbf{--}             & \textbf{--}/\textbf{--}             &  \textbf{--}/\textbf{--}      &-9.51/2.42       &  {\color{blue}-28.10} / {\color{blue}-39.44} & {\color{red}-60.02} /{\color{red}-41.98} \\ \hline
			Average  & \textbf{8.03/-5.02} & \textbf{-2.83/-26.25}   & \textbf{-7.92/-16.71} & \textbf{-5.31/-35.83} & \textbf{-4.47/-35.94}  & \textbf{-20.84/-34.84} & \textbf{-17.34/-43.96} & \textbf{{\color{blue}-29.4} /-14.14}& \textbf{-22.60/ {\color{blue}-46.66}} & \textbf{{\color{red}-39.39}  /{\color{red}-54.92}} \\ \bottomrule
			
		\end{tabular}
	}
	\caption{Compression on BDBR results calculated by the PSNR vs. bpp and MS-SSIM vs. bpp curves. H.265 is set as the benchmark. In all cases, our JCEVC achieves the best or 2nd best BDBR performance. On average, it significantly outperforms the state-of-the-arts.}
	\label{tab:bdbr}
\end{table*}

\section{Experiments}
\label{sec:exp}

\subsection{Experimental setup}
\label{subsec:settings}

\textbf{Datasets.} We train our JCEVC codec with the popular Vimeo-90k \cite{Vimeo90k} dataset, which consists of 89,000 video clips at a resolution of 448$\times$256. To report the performance of our method, we test on H.265 CTC (including Class B at 1920$\times$1080, Class C at 832$\times$480 and Class D at 416$\times$240) \cite{CTC}, MCL-JCV (at 1920$\times$1080) \cite{MCLJCV}, UVG (at 1920$\times$1080) \cite{UVG} and VTL (at 352$\times$288) \cite{VTL}. In total, there are 42 HD videos and 23 low-resolution videos are tested.

\textbf{Evaluation.} We compare our method with the popular deep codecs FVC \cite{HuZ:CVPR21}, Liu's \cite{LiuB:CVPR21}, RLVC \cite{YangR:JSTSP21}, Agustsson's \cite{Agustsson:CVPR20}, HLVC \cite{YangR:CVPR20}, M-LVC \cite{LinJ:CVPR20}, Hu's \cite{HuZ:ECCV20}, Lu's \cite{LuG:ECCV20}, DVC \cite{LuG:CVPR19} as well as H.265 implemented by x265 LDP very fast mode. For fair comparison, the results of compared methods are collected from their reports. The consumed bits and reconstruction quality are evaluated by bpp and PSNR/MS-SSIM, respectively. We also calculate the BDBR values \cite{BDBR} that represents the average bit reduction with the same PSNR or MS-SSIM.

\textbf{Implementation details.} We implement our model on Tensorflow with all training and testing performed on an NVIDIA RTX 2080Ti GPU. The batch size and $\gamma$ are set as 4 and 1000, respectively. For PSNR-oriented compression, we train four models with different $\lambda$ values from 512 to 2560; while for MS-SSIM-oriented compression, we train another four models with $\lambda$ from 8 to 48. Detailed training process can be seen in Section \ref{subsec:training}.

\subsection{Experimental results}
\label{subsec:results}

Figure \ref{fig:curves} shows the comparison between our JCEVC and the state-of-the-arts. To evaluate our method to the maximum extent, we test 6 video groups from 4 datasets and collect all available results of compared codecs. The performances of codecs are shown by two types of curves: PSNR vs. bpp and MS-SSIM vs. bpp. A curve above others is considered with a superior RD performance. Three conclusions can be drawn from the figure.
\textbf{First}, all deep codecs achieve competitive or superior performances compared with H.265 (x265 LDP very fast), which demonstrates the effectiveness of learning-based video coding.
\textbf{Second}, the deep video coding has been greatly improved since 2019, which demonstrates the potential of learning-based video coding. The recent deep codecs, such as FVC and Liu's, have significantly surpassed the H.265. We can envision a deep video codec with comparable performance to H.266/VVC in the foreseeable future.
\textbf{Third}, our JCEVC achieves significantly superior performances than the state-of-the-arts in most datasets. For example, in Class D by PSNR, UVG by MS-SSIM and VTL by PSNR, the JCEVC achieves remarkably higher performance even compared with the 2nd best curves. In Class C by PSNR and MS-SSIM, the JCEVC achieves comparable performances to FVC. While in other figures, the JCEVC also surpasses all compared methods. These facts undoubtedly demonstrate the superiority of our JCEVC model.

To quantitatively compare the video codecs, we also present the BDBR results (with PSNR and MS-SSIM) of all available curves in Table \ref{tab:bdbr}, where H.265 is also set as the benchmark. These results are consistent with those in Figure \ref{fig:curves} that our JCEVC always ranks the best or 2nd best in all datasets. An interesting result occurs when comparing FVC with JCEVC in MCL-JCV by MS-SSIM. The RD curves indicates JCEVC is superior while the BDBR slightly prefers the FVC. This conflict is due to the different definition domains to interpolate and calculate BDBR \cite{BDBR}, which is unusual and does not affect the conclusion. On average, the JCEVC achieves a BDBR of -39.39\% or -54.92\% by PSNR or MS-SSIM. This fact also supports the superior efficiency of the JCEVC.

By summarizing Figure \ref{fig:curves} and Table \ref{tab:bdbr}, there are some minor issues to be clarified. \textbf{First}, all codecs show weak improvements in CTC Class C. This fact might be attributed to higher motions in this Class \cite{LuG:ECCV20}. In particular, the temporal information (TI) values of Classes B, C, D are 18.6, 24.0 and 21.5, respectively. Despite that, our JCEVC still ranks the best and 2nd best in terms of BDBR by PSNR and MS-SSIM, respectively. \textbf{Second}, fair comparison with visual enhancement. This module was also adopted in HLVC. As shown above, our JCEVC is significantly superior to this method in terms of BDBR. \textbf{Third}, fair comparison with bi-directional IPPP structure. This structure was also used by RLVC while the hierarchical B structure was introduced by HLVC. Our JCEVC significantly outperforms the above two methods in terms of BDBR.

Regarding to computational complexity, some existing codecs did not report their time costs. Among all available codecs (DVC \cite{LuG:CVPR19}, Lu's \cite{LuG:ECCV20}, M-LVC \cite{LinJ:CVPR20}, RLVC \cite{YangR:JSTSP21}, FVC \cite{HuZ:CVPR21} and JCEVC, where the RLVC includes entropy coding for fair comparison), the time cost magnitudes are in the order of 1e-2s to 1s per frame. Our JCEVC is with a medium complexity of 0.246s (resp. 0.224s) per frame to encode (resp. decode) Class D on 1080Ti. This time cost is acceptable considering its promisingly high RD improvement compared with the other end-to-end deep video codecs.

\subsection{Ablation study}
\label{subsec:ablation}

Due to lack of space, the ablation experiments are conducted with PSNR-based JCEVC only. Similar conclusions can be drawn in terms of MS-SSIM vs. Bpp curves.

\begin{figure}[htbp]
	\centering
	\includegraphics[width=0.85\linewidth]{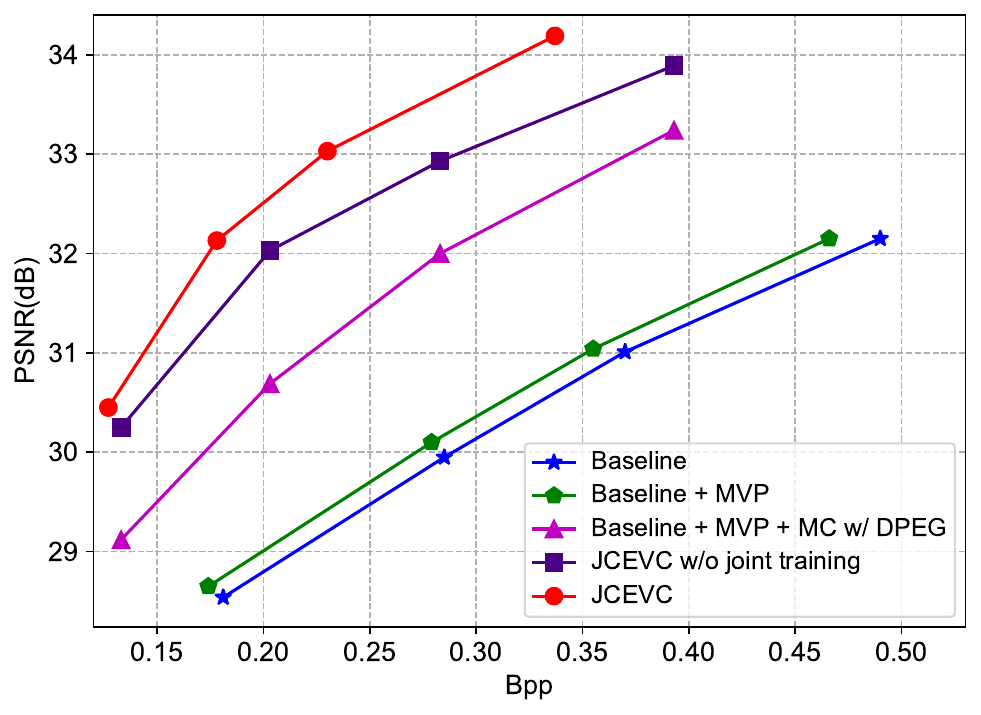}
	\caption{Contributions of all JCEVC modules. The RD performances are kept improved with more modules, which demonstrates the effectiveness of our design.}
	\label{fig:ablation}
\end{figure}

\begin{figure}[htbp]
	\centering
	\includegraphics[width=0.95\linewidth]{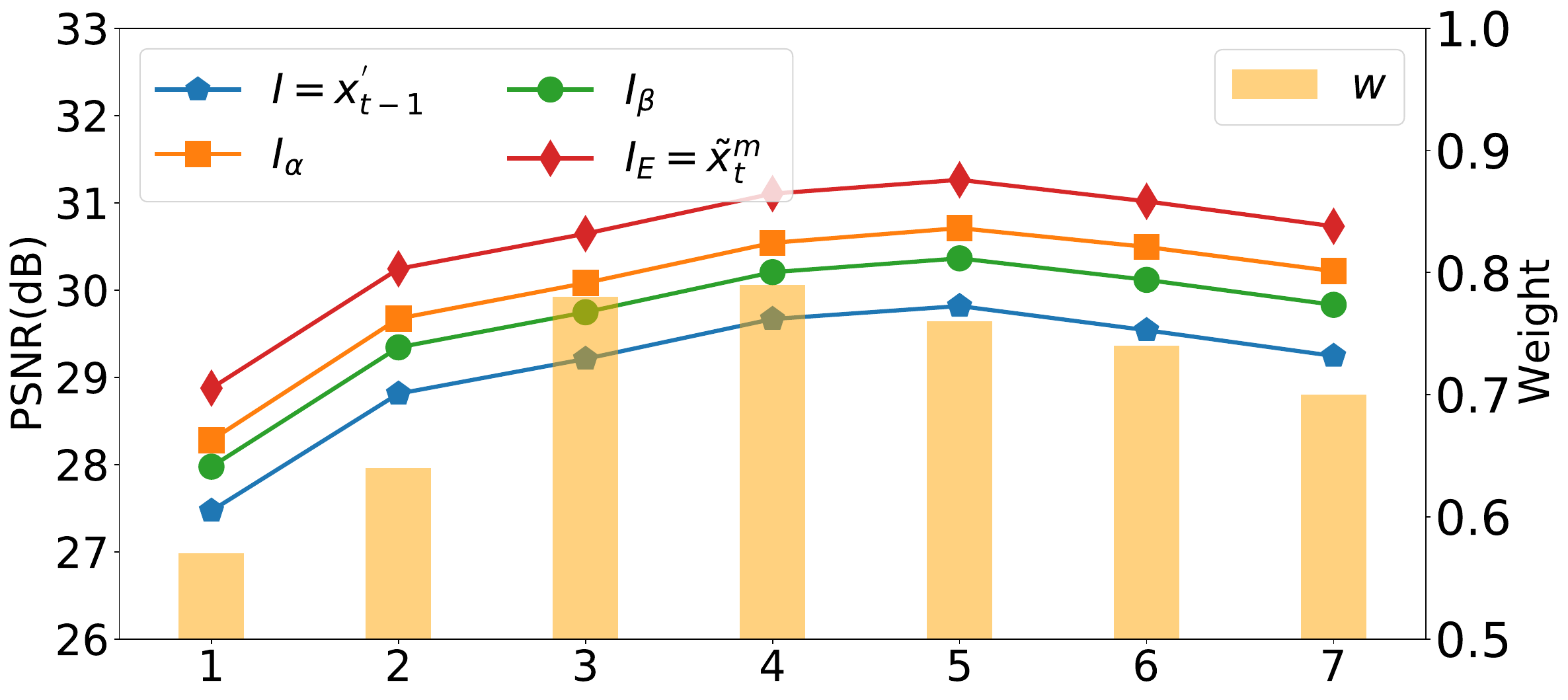}
	\caption{Comparison between the input ($I$), the intermediate results ($I_\alpha, I_\beta$) and the output ($I_E$) of DPEG in MC. Each path has its own contributions to the final output.}
	\label{fig:ablationmc}
\end{figure}

\textbf{Contributions of all modules.}  In JCEVC, we design an MVP with a light network, an MC with DPEG network, a PQE with DPEG networks and a joint training, as shown in Figure \ref{fig:framework}. To examine the contributions of these modules, we perform the following ablation study. A baseline method with a simple but feasible framework (ME+MVDC+MC w/o DPEG+RC) is examined first. Then, our designed modules (MVP, MC w/ DPEG, PQE, joint training) are introduced sequentially to observe their RD improvements. The average results on CTC class D are summarized in Figure \ref{fig:ablation}. With more designed modules, the RD performance is continuously improved. For example, with PSNR=32dB, the bpps of the five settings are 0.47, 0.45, 0.28, 0.20, 0.17, respectively, which indicate the bpp savings at 4.3\%, 37.8\%, 28.6\% and 15.0\% by introducing MVP, MC w/ DPEG, PQE and joint training. This fact reveals the effectiveness of our design, especially for the DPEG-based MC/PQE and the joint training.

\textbf{Contributions of individual paths in DPEG.} The DPEG network consists of an $\alpha$-path and a $\beta$-path. To observe the contributions of each path, we compare the input ($I = x_{t-1}^{'}$), the intermediate results ($I_\alpha, I_\beta$) and the output ($I_E = \tilde x_t^m$) of DPEG in MC. Figure \ref{fig:ablationmc} presents the results obtained by averaging the 1$\sim$7-th frames of all Class D sequences. It can be seen that both $I_\alpha$ and $I_\beta$ improve the PSNR of $I$. By fusing the results of $I_\alpha$ and $I_\beta$ with the weight $w$ ($w>0.5$ when $I_\alpha$ has a better visual quality), the resulted $I_E$ achieve a further high performance, which implies the complementarity between the $\alpha$-path and $\beta$-path as well as the effectiveness of our weight fusion.

\begin{figure}[htbp]
	\centering
	\includegraphics[width=0.85\linewidth]{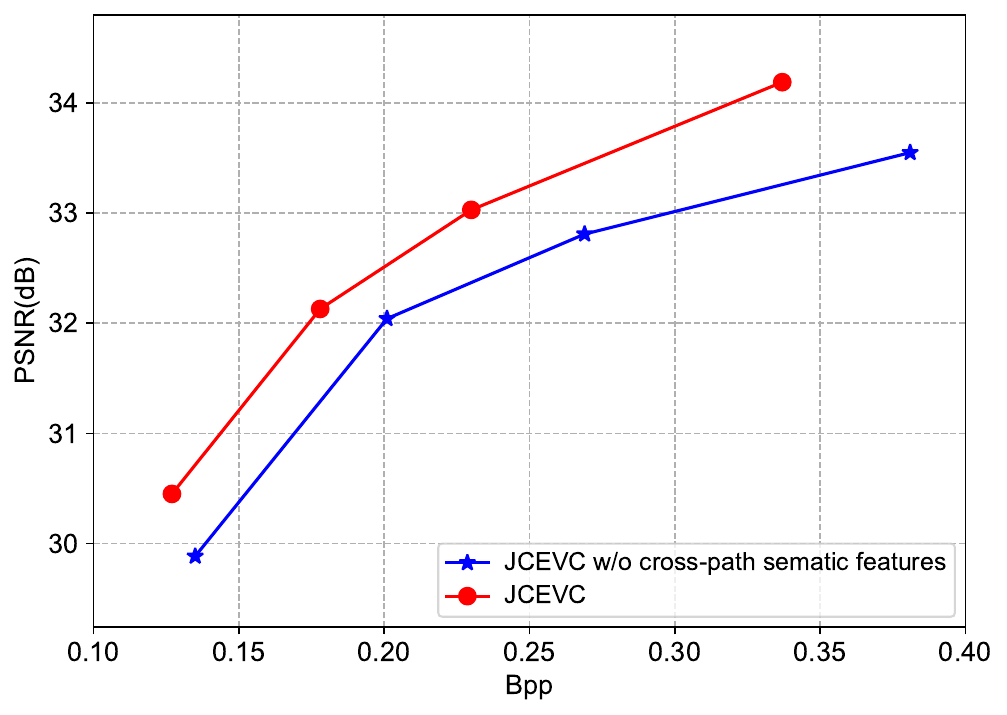}
	\caption{Comparison between JCEVC implementations with and without the cross-path sematic feature embedding. Our JCEVC improves its RD performance by introducing the cross-path sematic features. }
	\label{fig:ablationcrosspath}
\end{figure}

\begin{figure}[htbp]
	\centering
	\includegraphics[width=0.95\linewidth]{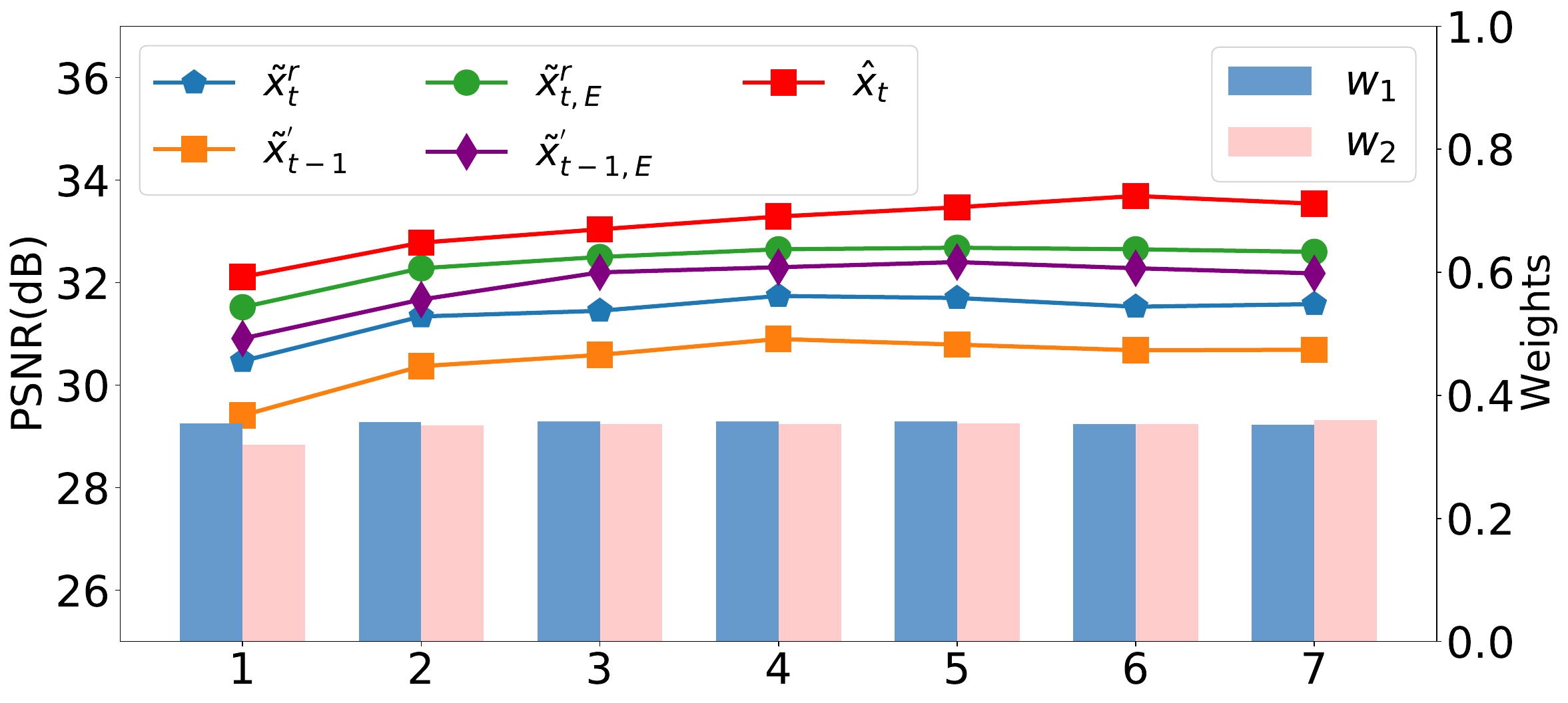}
	\caption{Comparison between the inputs ($\tilde x_t^r, \tilde x_{t-1}^{'}$), the intermediate results ($\tilde x_{t,E}^r, \tilde x_{t-1,E}^{'}$) and the output ($\hat x_t$) in PQE. Both inputs are enhanced and further fused to generate the reconstructed frame $\hat x_t$ that is of the highest quality among all these results.}
	\label{fig:ablationpqe}
\end{figure}

\textbf{Contributions of cross-path sematic feature embedding.} In Figure \ref{fig:dpeg}, we design the DPEG network with the sematic features $F_{t,1} \in R^{H \times W \times C}, F_{t,2} \in R^{\frac{H}{2} \times \frac{W}{2} \times C}, F_{t,3} \in R^{\frac{H}{4} \times \frac{W}{4} \times 2C}$ that are fed from $\alpha$-path to $\beta$-path. To investigate the impact of this cross-path sematic feature embedding, we compare our JCEVC encoder with its reduced version without these cross-path sematic features and summarize the RD performances in Figure \ref{fig:ablationcrosspath}, where the results are obtained by averaging the RD performances of all Class D sequences. Compared with x265 LDP very fast, the BDBR values of the two settings are -34.26\% and -44.72\%, respectively. It can be easily concluded that the cross-path features $F_{t,1}, F_{t,2}, F_{t,3}$ contributes to the final coding performance.

\textbf{Contributions of DPEG networks in PQE.} The PQE module utilizes two DPEG networks to enhance $\tilde x_t^r$ and $\tilde x_{t-1}^{'}$ as $\tilde x_{t,E}^r$ and $\tilde x_{t-1,E}^{'}$, respectively. Then, it applies a weighted sum of $\tilde x_t^r$, $\tilde x_{t,E}^r$ and $\tilde x_{t-1,E}^{'}$ to obtain the final reconstruction $\hat x_t$. Figure \ref{fig:ablationpqe} shows the results of these pictures by averaging the 1$\sim$7-th frames of all Class D sequences. Obviously, each DPEG network contributes to the final performance. With a weighted fusion, the finally reconstructed frame $\hat x_t$ is of a high visual quality in terms of PSNR. Therefore, it is reasonable to apply two DPEG networks in the PQE module of JCEVC.

\begin{figure}[htbp]
	\centering
	\includegraphics[width=0.85\linewidth]{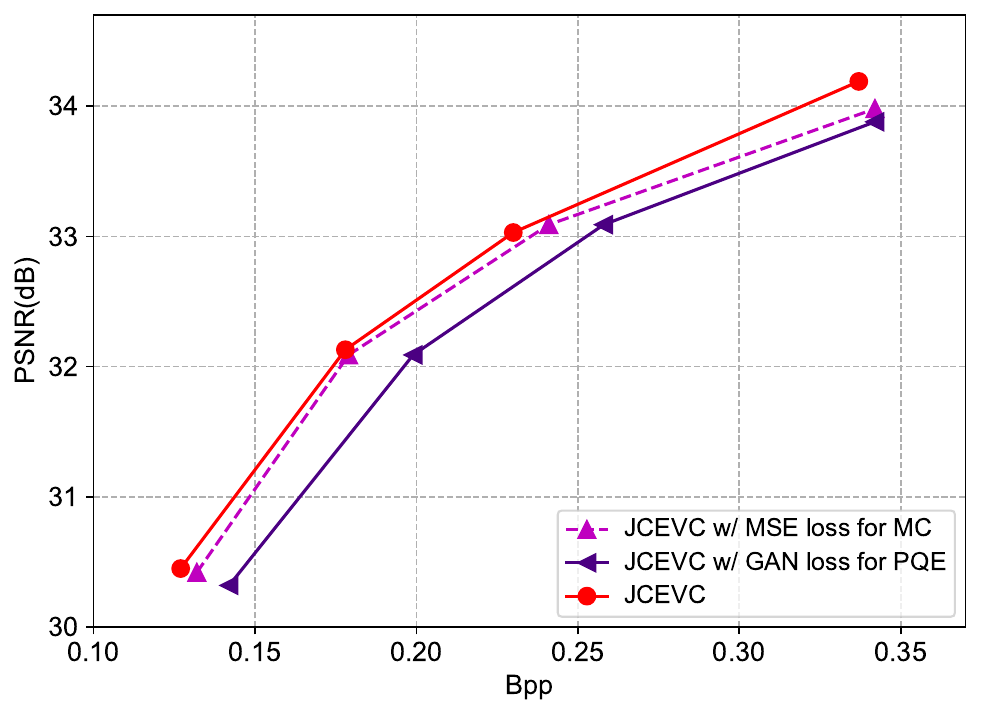}
	\caption{Comparison between JCEVC implementations with different loss functions. The RD performances of JCEVC is decreased with MSE loss for MC or GAN loss for PQE.}
	\label{fig:ablationloss}
\end{figure}

\textbf{Effectiveness of loss functions.} The JCEVC adopts GAN loss and MSE loss in MC and PQE modules, respectively. With different loss functions, {\it e.g.}, MSE loss for MC, or GAN loss for PQE, the JCEVC achieves inferior RD performances, which can be seen in Figure \ref{fig:ablationloss}. As discussed in Section 3.4, the joint training allows a higher $D$ in MC and further minimize it in PQE module. These different distortion constraints might be located in the attainable and unattainable regions of the perception-distortion tradeoff \cite{BlauY:CVPR18}, respectively. This can be taken as a plausible explanation that we should use different loss functions in different modules.

\subsection{Limitations}
\label{subsec:limit}

The H.266/VVC incorporates the enhanced block partitioning, diversified intra and inter predictions, refined ME and MC, extended transform and quantization, improved entropy coding and adaptive deblocking filters with RD optimization \cite{Bross:TCSVT21}. As a contrast, the deep video codecs, including our JCEVC and all compared methods \cite{Agustsson:CVPR20, HuZ:ECCV20, HuZ:CVPR21, LinJ:CVPR20, LiuB:CVPR21, LuG:CVPR19, LuG:ECCV20, YangR:CVPR20, YangR:JSTSP21}, are still imitating the traditional coding modules, which limits their RD performances compared with the newest video coding standard. In such sense, the development of more deep coding methods are imperative to realize more advanced video coding techniques. A hybrid framework to take advantages of all deep codecs is feasible. In addition, the deep video codecs prevail in the utilization of big video data. With a joint training of all modules on large-scale datasets, the deep video coding has a brighter outlook in a foreseeable future.

\section{Conclusions}
\label{sec:end}

Nowadays, the deep video codecs have been extensively studied with ever-increasing RD performance. To compete with reigning codecs, a
high-efficiency deep codec is strongly desired. In this paper, we proposed an end-to-end deep video codec called JCEVC that consists of ME, MVP, MVDC, MC, RC and PQE modules. We designed a DEPG network with dual-path generators and cross-path sematic feature embedding, and further reused it in both MC and PQE modules. Aiming at a global optimization of the RD performance, we also employed a joint training of deep video compression and enhancement. Comprehensive studies on four popular datasets have demonstrated the RD efficiency of our JCEVC method, which outperforms the state-of-the-art deep video codecs.

{\em Sourcecode of this paper will be released after the peer-review process.}

\clearpage

\bibliographystyle{ACM-Reference-Format}
\bibliography{egbib}
\end{document}